\documentclass[twocolumn]{aastex631}
\shorttitle{Variations of the IMF with the environment}
\shortauthors{Dib}
\graphicspath{{./}{figures/}}
\begin{document}

\title{Variation of the high-mass slope of the stellar initial mass function: Theory meets observations}

%% you desire. In v6.31 you can include a footnote in the title.

\correspondingauthor{Sami Dib}
\email{sami.dib@gmail.com, dib@mpia.de}

\author[0000-0002-8697-9808]{Sami Dib}
\affiliation{Max Planck Institute for Astronomy, K\"{o}nigstuhl 17, 69117, Heidelberg, Germany}

\begin{abstract}
We present observational evidence of the correlation between the high-mass slope of the stellar initial mass function (IMF) in young star clusters and their stellar surface density, $\sigma_{*}$. When the high-mass end of the IMF is described by a power law of the form $dN/d{\rm log}{M_{*}}\propto M_{*}^{-\Gamma}$, the value of $\Gamma$ is seen to weakly decrease with increasing $\sigma_{*}$, following a $\Gamma=1.31~\sigma_{*}^{-0.095}$ relation. We also present a model that can explain these observations. The model is based on the idea that the coalescence of protostellar cores in a protocluster forming clump is more efficient in high density environments where cores are more closely packed. The efficiency of the coalescence process is calculated as a function of the parental clump properties and in particular the relation between its mass and radius as well as its core formation efficiency. The main result of this model is that the increased efficiency of the coalescence process leads to shallower slopes of the IMF in agreement with the observations of young clusters, and the observations are best reproduced with compact protocluster forming clumps. These results have significant implications for the shape of the IMF in different Galactic and extragalactic environments and have very important consequences for galactic evolution.
\end{abstract}


\section{Introduction} \label{sec:intro}

The stellar initial mass function (IMF) which is the distribution of the masses of stars at their birth is of fundamental importance to almost all branches in astrophysics. The shape of the IMF affects the efficiency of star formation in molecular clouds (Dib 2011; Dib et al. 2011,2013; Hony et al. 2015; Wells et al. 2022; Gardu\~{n}o et al. 2023), and sets the amount of feedback from stars into the large scale interstellar medium (Dib et al. 2006,2009,2021; Gatto et al. 2017, Orr et al. 2020; Lara-L\'{o}pez et al. 2023). Furthermore, the IMF, and particularly its component at the high-mass end determines the chemical and dynamical evolution of galaxies (Prantzos \& Boissier 2000; Dav\'{e} et al. 2011; Brusadin et al. 2013; Few et al. 2014; Recchi \& Kroupa 2015; Dariush et al. 2016; Matteucci 2016; Vincenzo et al. 2016; Fontanot et al. 2017; Diehl \& Prantzos 2023). In the nearby Galactic field, Salpeter (1955) found that the present-day stellar mass function can be described by a power law $\left(dN/d{\rm log}M_{*}\right)=M_{*}^{-\Gamma}$, where $dN$ is the number of stars between ${\rm log}M_{*}$ and ${\rm log} M_{*}+d{\rm log}M_{*}$, with $\Gamma \approx 1.35$. Subsequent surveys of larger volumes of the Milky Way enabled a more complete characterization of the shape of the present-day stellar mass function down to the low mass and substellar mass regimes. The results from these surveys suggest that the mass function of stars in the Galactic field rises from the brown dwarf and low stellar mass regime until it peaks at $\approx 0.3$ M$_{\odot}$, after which it declines steeply in the intermediate-to-high mass regime (Miller \& Scalo 1979; Scalo 1986; Kroupa 2001, Chabrier 2003; Bochanski et al. 2010; Parravano et al. 2011; Sollima et al. 2019). Mor et al. (2019) combined the constraints from the stars chemical composition and from their dynamics and spatial distribution and relaxed the assumption of a constant Galactic star formation rate (SFR) made in earlier studies. By matching the observational data to the Besancon model of the Galaxy (Robin et al. 2003; Mor et al. 2018), they found that the time averaged Galactic IMF is shallower at both the low and high-mass ends than what was previously found. This result has been predicted by Dib \& Basu (2018) who attributed it to the existence of cluster-to-cluster variations of the IMF in Galactic clusters (see also Dib 2022). Variations of the IMF between galaxies have also been reported (Gunawardhana et al. 2011, Mart\'{i}n-Navarro et al. 2015, Geha et al. 2013, Gennaro et al. 2018), and these have been successfully interpreted as being due to variations in the average galactic gas-phase metallicity and SFR, and to the existence of various levels of IMF variations in their population of clusters (Yan et al. 2017, Je\v{r}\'{a}bkov\'{a} et al. 2018, Dib et al. 2022). 

For individual stellar clusters, measurements of the IMF, both in the Milky Way and in external nearby galaxies have led to discrepant conclusions in terms of their similarity with the Galactic field stellar mass function. Earlier studies argued that the observed level of variations between the IMFs of young Galactic clusters, particularly at the high-mass end, are compatible with a universal value within the 1$\sigma$ uncertainties (Bastian et al. 2010; Offner et al. 2014). However, recent studies suggest that there are non-negligible cluster-to-cluster variations in the set of parameters that characterize the shape of the IMF among the population of young Galactic stellar clusters (Sharma et al. 2008; Hsu et al. 2012; Dib 2014; Mallick et al. 2014; Lim et al. 2015a; Kraus et al. 2017; Dib et al. 2017; Zhang et al. 2018; Rangwal et al. 2023; Kaur et al. 2023; see also Schneider et al. 2018 for the case of the 30 Dor starburst region in the Large Magellanic Cloud). In particular, Dib et al. (2017) showed that the fraction of single O stars measured in a large sample of young Galactic clusters (341 clusters from the Milky Way Stellar Clusters Survey; Kharchenko et al. 2013; Dib et al. 2018) can only be reproduced by populations of Galactic clusters that have a significant intrinsic scatter in the set of parameters that characterize their IMFs. They found that the standard deviation of the distribution of $\Gamma$ values is $\sigma_{\Gamma} \approx 0.6$. In M31, the results of Weisz et al. (2015) also hint to the existence of cluster-to-cluster variations of the slope of the IMF in the intermediate to high-mass regime. There are also indications that the shape of the IMF of massive clusters that are the progenitors to the preset day globular clusters possess some degree of variations, both in their characteristic mass and slope at the high-mass end (Wang et al. 2020, Kravtsov et al. 2022; Wirth et al. 2022; Dib et al. 2022) 

\begin{figure}
\begin{center}
\includegraphics[width=\columnwidth] {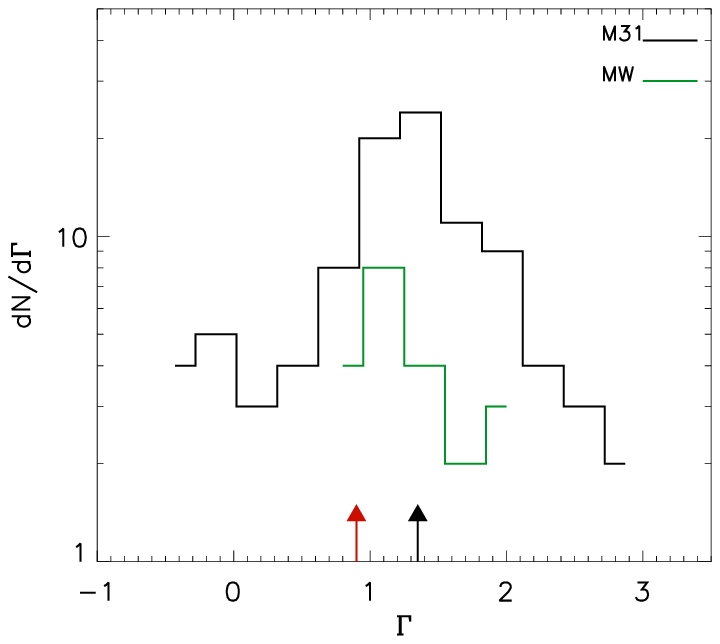}
\caption{Distributions of the high-mass slopes for the sample of young Galactic clusters measured by Lim et al. (2015a) and of clusters in M31 constructed using data from Weisz et al. (2015). In Lim et al. (2015a) and Weisz et al. (2015), the values of $\Gamma$ have been measured using stars that are more massive that 1 M$_{\odot}$ and 2 M$_{\odot}$, respectively. The black arrow marks the position of the Salpeter value of $\Gamma=1.35$ for the present-day stellar mass function of the Milky Way and the red arrow marks the value of $\Gamma \approx 0.9$ found by Mor et al. (2019) for the time-averaged galaxy integrated initial stellar mass function of the Milky Way.}
\label{fig1}
\end{center}
\end{figure}

When attempting to investigate whether the slope of the IMF in the intermediate to high-mass range truly varies, it is necessary to compare clusters whose IMF has been derived using the same methodology, that is, where photometric or spectroscopic measurements have been derived using the same calibrations and instruments and that have accounted for the effects of reddening and cluster membership using the same procedures. Such homogenous samples of clusters for which the IMF has been derived in a similar fashion are rare. Lim et al. (2015a) used data from the Sejong open cluster survey (SOS, Sung et al. 2013a) and derived the slope of the IMF at the high-mass end for a sample of 16 Galactic young clusters (median age of their stellar populations are in the range $\approx 0.8$ to $5$ Myrs)\footnote{Lim et al. (2015a) summarizes the findings of several papers that were already published by members of the SOS surveys (Sung \& Bessell 2004,2010; Hur et al. 2012; Lim et al. 2013; Sung et al. 2013b; Lim et al. 2014a,b; Lim et al. 2015b,c; Hur et al. 2015) and others which appeared later (Sung et al. 2017; Kim et al. 2021).}, whereas Weisz et al. (2015) derived the same quantity for a sample of 85 young clusters in M31. Figure~\ref{fig1} displays the distribution of the measured slopes for both samples. The mean values of $\Gamma$ are $1.33$ and $1.37$ for the Milky Way and the M31 sample and the standard deviations are $\sigma_{\Gamma}=0.36$ and $0.68$ for the Milky Way and M31 samples, respectively. Both standard deviations are of the order of, or larger than the mean individual uncertainties for clusters in each sample. This level of scatter in $\Gamma$ is comparable to the standard deviation of $\sigma_{\Gamma}\approx 0.6$ found by Dib (2014) for a smaller number of Galactic young clusters.

The aim of this paper is to review some of the evidence for IMF variations in young clusters, particularly in terms of the slope at the high-mass end and to explain the origin of such variations. In \S.~\ref{review}, we present a compilation of observational data which indicates a correlation between the value of $\Gamma$ in young clusters on their stellar surface density, $\sigma_{*}$. In \S.~\ref{gravturb}, we discuss the limitations of theoretical models based on the gravoturbulent fragmentation of molecular clouds in terms of predicting variations in the value of $\Gamma$ and we review some of the other existing ideas, such as those based on gas accretion by protostellar cores as well as the effects of the coalescence of cores. In \S.~\ref{model}, we introduce the different components of a model of core coalescence in protocluster forming clumps. Using this model, we describe in \S.~\ref{coevol}, the co-evolution of the mass function of dense cores and the IMF of the newly formed stars in the clumps. In \S.~\ref{paramstudy}, we perform a large parameter study and compare a series of models to the observations in terms of the relation between $\Gamma$ and $\sigma_{*}$. In \S.~\ref{conc}, we summarize our findings and conclude. 

\section{Observational evidence for IMF variations with the environment}\label{review}

\begin{figure}
\begin{center}
\includegraphics[width=\columnwidth] {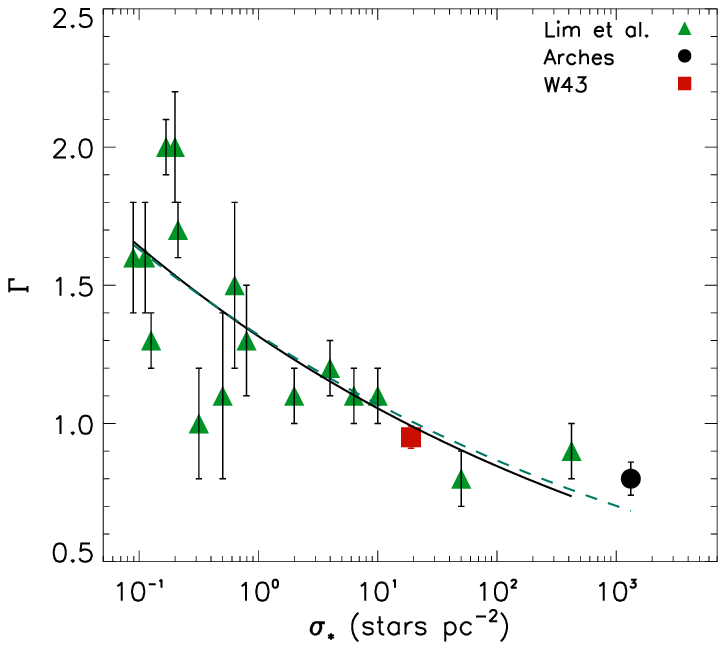}
\caption{The slope of the IMF in the intermediate to high-mass end plotted as a function of the stellar surface density, $\sigma_{*}$. The green triangles correspond to the sample of young Galactic clusters of Lim et al. (2015a). The black circle corresponds to the Arches cluster. The red point corresponds to the massive Galactic star forming region W43. The full black line is a power law fit to the Lim et al. data alone whereas the green dashed line is a power law fit to the entire data set (see text for detail).}
\label{fig2}
\end{center}
\end{figure}

Equally interesting to uncovering variations of the IMF is to understand the origin of such variations. Lim et al. (2015a) made estimates of the stellar surface density ($\sigma_{*}$) in the clusters they have observed for stars more massive that $5$ M$_{\odot}$. Figure \ref{fig2} displays the values of $\Gamma$ versus $\sigma_{*}$ of the Lim et al. sample. We complement this sample with the Arches cluster which is known to be the densest and most massive cluster in the Milky Way (Espinoza et al. 2009; Harfst et al. 2010; Clarkson et al. 2012). Based on its most recent census, the Arches cluster contains $636$ stars with masses $\gtrsim 2.5$ M$_{\odot}$ and the cluster half-light radius is $\approx 0.48$ pc (Hosek et al. 2015). Using these values yields a stellar surface density of $\approx 1325$ stars pc$^{-2}$. The number could be smaller if we were counting stars more massive than 5 M$_{\odot}$, as in Lim et al. (2015a). For the $\Gamma$ value of the Arches cluster, we also adopt the latest estimate made by Hosek et al. (2019) who derived a value of $0.8\pm0.05$. This value is similar to earlier estimates made by Stolte et al. (2002), Kim et al. (2006), and Habibi et al. (2013). Also shown in Fig.~\ref{fig2} is a data point corresponding to the Galactic massive star forming region W43. Pouteau et al. (2022) derived the value of $\Gamma$ for W43 by fitting the mass function of dense cores (hereafter, the CMF) with a power law. They obtained a value of $\Gamma=0.95$ and a core surface density of $\approx 19$ cores pc$^{-2}$. 

 In Fig.~\ref{fig2}, a dependence of $\Gamma$ on $\sigma_{*}$ is observed, with denser clusters having shallower IMFs. In the low stellar surface density regime ($\lesssim 1$ stars pc$^{-2}$), the scatter is relatively large and corresponds to the range of values reported by Dib et al. (2017). On the other hand, there are no known high surface density young clusters in the Milky Way that possess high values of $\Gamma$. A power-law fit to the data points of Lim et al. alone yields

\begin{equation}
\Gamma=1.31(\pm0.06) \sigma_{*}^{-0.095\pm0.026}.
\label{eq1}
\end{equation}

The fit is overlaid to the data in Fig.~\ref{fig2} (full black line). We also fit another power-law combining the Lim et al. sample and the data points for Arches and W43. In this case, a power law fit yields

\begin{equation}
\Gamma=1.32(\pm0.058) \sigma_{*}^{-0.091\pm0.020}.
\label{eq2}
\end{equation}

\noindent and the fit is also overlaid to the data in Fig.~\ref{fig2} (green dashed line). We adopt Eq.~\ref{eq1} as our fiducial fit to the data as it is based on a homogenous sample of clusters, but both Fig.~\ref{fig2} and Eq.~\ref{eq2} seem to indicate that this derivation is robust as the inclusion of other clusters does not seem to modify Eq.~\ref{eq1} substantially.

The results displayed in Fig.~\ref{fig2} resemble, qualitatively, those presented by Marks et al. (2012) who found a relationship between the slope at the high-mass end and the volume density of the clusters' progenitor clouds. In Marks \& Kroupa (2012), the cloud densities were derived by matching the high-mass slope and total stellar mass of each cluster (mostly globular clusters) with a grid of N-Body models that contains parametrized prescriptions for the star formation efficiency and gas expulsion timescales (Marks \& Kroupa 2010). While the dependence between the value of $\Gamma$ and the cloud density or surface density is qualitatively the same between this work and Marks et al. (2012), our results, in addition to being derived directly from the observations, they also seem to suggest that such a dependence between $\Gamma$ and the stellar surface density extends to lower clusters masses which is a regime that was not explored in Marks et al. (2012). 

\section{Limitations of current theoretical models} \label{gravturb}

In contemporary models of star formation that are based on the idea of gravoturbulent fragmentation, the interplay between gravity, supersonic turbulence, and magnetic fields is believed to be responsible for the formation of dense structures within molecular clouds and clumps (Klessen 1997; Dib et al. 2008a,2010a; Bate 2009; Federrath et al. 2014; Mathew et al. 2023). In analytical formulations of this gravoturbulent scenario, once the cores have formed, the effects of gravity are only taken into account in terms of the self-gravity of the cores, and they are balanced by the dispersive forces (Padoan \& Nordlund 2002; Hennebelle \& Chabrier 2008; Hopkins 2012). In these models, the resulting shape of the mass function of dense cores at the high-mass end can be described by a power law function whose exponent is directly related to the exponent of the power law that describes the velocity field power spectrum. Because the properties of turbulence within molecular clouds are observed to be quasi universal (Heyer \& Brunt 2004), it is unlikely that variations at the high-mass end in the slope of the CMF and consequently of the IMF can result from the effects of turbulence alone.

Gravoturbulent fragmentation models neglect the long range effects of gravity that can be either the accretion of gas by the cores, or the effects of core coalescence. The effects of gas accretion by protostellar cores has been investigated in some detail by several authors (Zinnecker 1982; Smith 1985; Bonnell et al. 2001; Basu \& Jones 2004; Myers 2008; Dib et al. 2010b; Veltchev et al. 2011; Maschberger 2013; Hoffman et al. 2018; Clark \& Whitworth 2021). In particular, Dib et al. (2010b) argued that depending on the accretion rates and accretion timescales, gas accretion can result in a modification of the initial core mass function (ICMF) inherited from gravoturbulent fragmentation, leading to the development of a shallower function in the high-mass regime and to the depletion of low mass cores by shifting them towards higher masses. The coalescence of protostellar cores has been primarily invoked to explain the shallower than Salpeter slopes of the IMF in massive starburst clusters (Nakano 1966; Silk \& Takahashi 1979; Podsiadlowski \& Price 1992; Price \& Podsiadlowski 1995; Murray \& Lin 1996; Shadmehri 2004; Dib et et al. 2007; Dib 2007,2008b; Huang et al. 2013). The model presented by Dib et al. (2007) describes the co-evolution of the CMF and IMF in dense environments where protostars are closely packed and where the coalescence process is efficient. This model successfully explains the occurrence of shallow slopes in the intermediate to high-mass regimes in massive starburst clusters such as the Arches cluster, NGC 3603, Westerlund 1, and the Quintuplet cluster (Stolte et al. 2006; Espinoza et al. 2009; Hu{\ss}mann et al. 2012; Lim et al. 2013; Hosek et al. 2019), as well as in ultra compact dwarf galaxies (Dabringhausen et al. 2009). The aim of this work is to investigate whether the process of core coalescence in protocluster forming clumps can provide a theoretical interpretation to the observed relation between $\Gamma$ and $\sigma_{*}$ as the one seen in Fig.~\ref{fig2}.

\section{Model}\label{model}

We describe below the components of a relatively simple model that follows the coalescence of dense cores in a protocluster forming clump. Cores are overdense structures that form within the clump, presumably by gravoturbulent fragmentation, with a prescribed ICMF. The model explores how the ICMF is modified by core coalescence under various conditions and what is the shape of the resulting IMF after cores have collapsed to form stars. 

\subsection{Clump properties}\label{clumpbasics}

We consider here that cores form in spherically symmetric protocluster clumps with a uniform gas density of mass $M_{\rm cl}$, and radius $R_{\rm cl}$, with a relationship between $M_{\rm cl}$ and $R_{\rm cl}$ that follows 

\begin{equation}
M_{\rm cl}=A_{\rm cl}\times R_{\rm cl}^{\beta},
\label{eq3}
\end{equation}

where $A_{\rm cl}$ is a scaling coefficient. The values of $\beta$ and $A_{\rm cl}$ are likely to depend on the environment, and from an observational perspective, on the gas density tracer used to measure gas masses, and possibly on aspects of the clump finding algorithms. In the Central Molecular Zone, Li \& Zhang (2020) found $M_{\rm cl} ({\rm M}_{\odot}) \approx 1000 R_{\rm cl}^{3}(\rm pc)$ using $870$ $\mu$m dust measurements and an assumed gas-to-dust ratio of $100$. In the Milky Way disk, Dib et al. (2010b) used C$^{18}$O ($J = 1-0$) line observations of molecular clumps from Saito et al. (2007) and obtained $M_{\rm cl} ({\rm M}_{\odot})=10^{3.62\pm0.14} R_{\rm cl}^{2.54\pm0.25}(\rm pc)$. On the other hand, Urquhart et al. (2014) found $\beta \approx 1.64$ and $A_{\rm cl} \approx 2600$ for the ensemble of clumps in the ATLASGAL survey. However, the most massive protocluster clouds in the Galaxy seem to lie above their derived mass-radius relation, indicating a larger normalization and larger value of $\beta$. In the Large Magellanic Cloud, Mok et al. (2021) found a value of $\beta=2.77$. As fiducial values, we adopt values that are very close to those found by Dib et al. (2010b) using the Saito et al. (2007) data, namely $\beta=2.5$ and $A_{\rm cl}=4000$. However, it is not expected that the entire coeval galactic population of protocluster clumps follow the same mass-radius relation. Therefore, we also consider values of $\beta=2.75$ and $3$, as well as values of $A_{\rm cl}=3000$ and $5500$. 

At any given time in the lifetime of the protocluster clump, only a fraction of the clump's mass would be converted into dense cores and there is a likely relationship between the clump size and mass and its star formation activity (Dib 2023). While cores can continue to form in the clump, for simplicity, we will assume in this work that all cores are formed at once with a given core formation efficiency (CFE), such that the total mass in cores is given by $M_{\rm c}={\rm CFE} \times M_{\rm cl}$. At first glance, it might appear that there is a degeneracy between the choice of $M_{\rm cl}$ and the CFE as some combinations of these two quantities can result in the same value of $M_{\rm c}$, and consequently in the same number of cores of a given mass in the initial core mass function (ICMF). However, the number of cores for any specific combination of $M_{\rm cl}$ and CFE will be uniquely related to a clump mass and size. We vary the value of the CFE between 0.05 and 0.25 and consider the case of clumps in the mass range of $5\times10^{3}$ M$_{\odot}$ to $4\times10^{4}$ M$_{\odot}$.

\subsection{Core properties and the initial core mass function}\label{corbasics}

\begin{figure}
\begin{center}
\includegraphics[width=0.9\columnwidth] {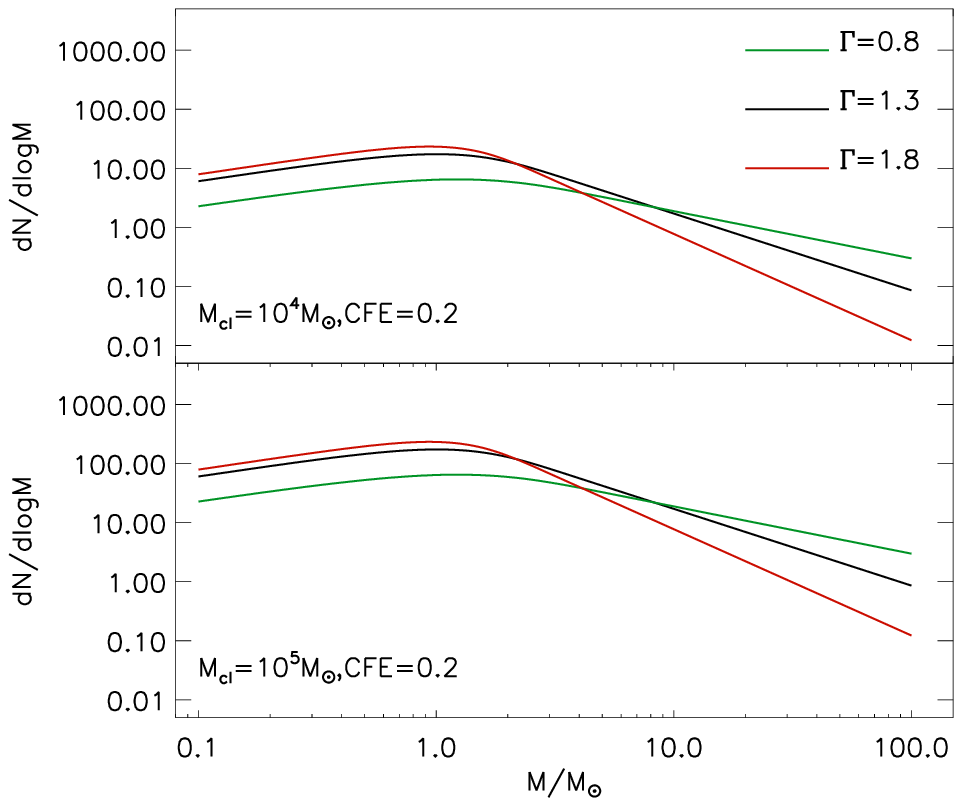}
\vskip 1cm
\caption{Examples of the ICMF for fixed values of $\gamma=0.6$ and $M_{\rm ch}=1.2$ M$_{\odot}$ and for various values of $\Gamma$. The upper sub-panel displays the case of a $10^{4}$ M$_{\odot}$ protocluster clump and the lower sub-panel a protocluster clump whose mass is $10^{5}$ M$_{\odot}$. In both cases, the CFE is equal to 0.2 leading to a total mass in cores of $2\times10^{3}$ M$_{\odot}$ and $2\times10^{4}$ M$_{\odot}$, respectively.}
\label{fig3}
\end{center}
\end{figure}

The rate at which cores coalesce in a protocluster clump depends on a certain number of physical quantities. These include the numbers of cores of any given mass present in the clump, the cross section of the cores, and the relative velocity between them. The initial distribution of the protostellar core masses are described by the ICMF. The shape of the ICMF, prior to the effects of core coalescence and gas accretion, is mostly set by the competition between the core's self-gravity, turbulence, and magnetic fields, and in some clumps by the incident external radiation field which can heat the star forming gas and affect its chemistry. When observed with low density tracers such as the CO (1-0) transition line, substructures in molecular clouds follows a power law with an exponent of $\approx 0.7$ where the ICMF is described by $dN/d{\rm log}(M) \propto M^{-\Gamma}$. However, such cores and clumps are mostly unbound. Dense cores within protocluster clumps follow a power with a larger exponent, typically in the range of $\Gamma \approx 1$ to $1.5$. While cores that lead to the formation of individual stars or binaries are gravitationally bound, cores that participate in the coalescence process in the clump do not need to be all gravitationally bound.  

In this work, we consider a simple parametric form of the ICMF which is characterized by two power laws in the low and high-mass regimes, and a characteristic mass (i.e., peak mass). Such a function is given by a tapered power law (De Marchi et al. 2010; Parravano et al. 2011; Dib et al. 2017)

\begin{equation}
\phi_{\rm i} (M)= \phi_{\rm 0}  M^{-\Gamma-1} \left\{1-\exp\left[-\left(\frac{M}{M_{\rm ch}}\right)^{\gamma+\Gamma}\right] \right\}\
\label{eq4}
\end{equation}

where $\phi_{\rm 0}$ is normalization constant, which is simply set by $M_{\rm c}=\phi_{\rm 0} \int_{M_{\rm l}}^{M_{\rm u}} M \phi_{\rm i}(M) dM$, where $M_{\rm l}$ and $M_{\rm u}$ are the lower and upper mass limits of cores in the ICMF and in all calculations are fixed to values of $0.1$ M$_{\odot}$ and 100 M$_{\odot}$, respectively. Figure \ref{fig3} displays examples of the ICMF for three selected values of $\Gamma=0.8, 1.3,$ and $1.8$. The other two parameters have fixed values of $\gamma=0.6$ and $M_{\rm ch}=1.2$ M$_{\odot}$. The normalization corresponds to clump masses of $10^{4}$ M$_{\odot}$ (upper sub-panel) and $10^{5}$ M$_{\odot}$ (lower sub-panel) with a CFE$=0.2$ in both cases. In the remaining of the text, we will consider as an input ICMF one that has fixed parameters of $\Gamma=1.3$, $\gamma=0.6$, and $M_{\rm ch}=1.2$ M$_{\odot}$.

\begin{figure}
\begin{center}
\includegraphics[width=0.9\columnwidth] {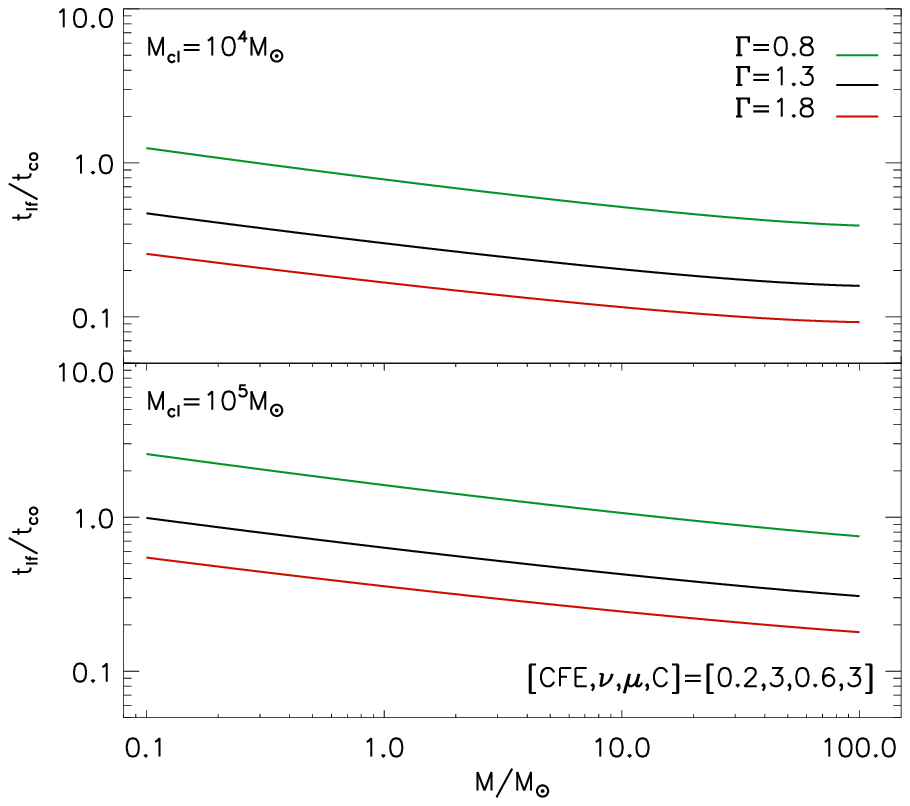}
\vskip  1cm
\caption{Ratios of the infall timescales to the coalescence timescales of cores for a core mass function with fixed values of $\gamma=0.6$ and $M_{\rm ch}=1.2$ M$_{\odot}$ and for various values of $\Gamma$, and for the fiducial values of the cores parameters. The upper panel displays the case of a $10^{4}$ M$_{\odot}$ protocluster clump and the lower panel corresponds to a protocluster clump whose mass is $10^{5}$ M$_{\odot}$. In both cases, the CFE is equal to 0.2 leading to a total mass in cores of $2\times10^{3}$ M$_{\odot}$ and $2\times10^{4}$ M$_{\odot}$, respectively. The parameters of the mass-radius scaling relation are $\beta=2.5$ and $A_{\rm cl}=4000$.}
\label{fig4}
\end{center}
\end{figure}

\begin{figure}
\begin{center}
\includegraphics[width=0.9\columnwidth] {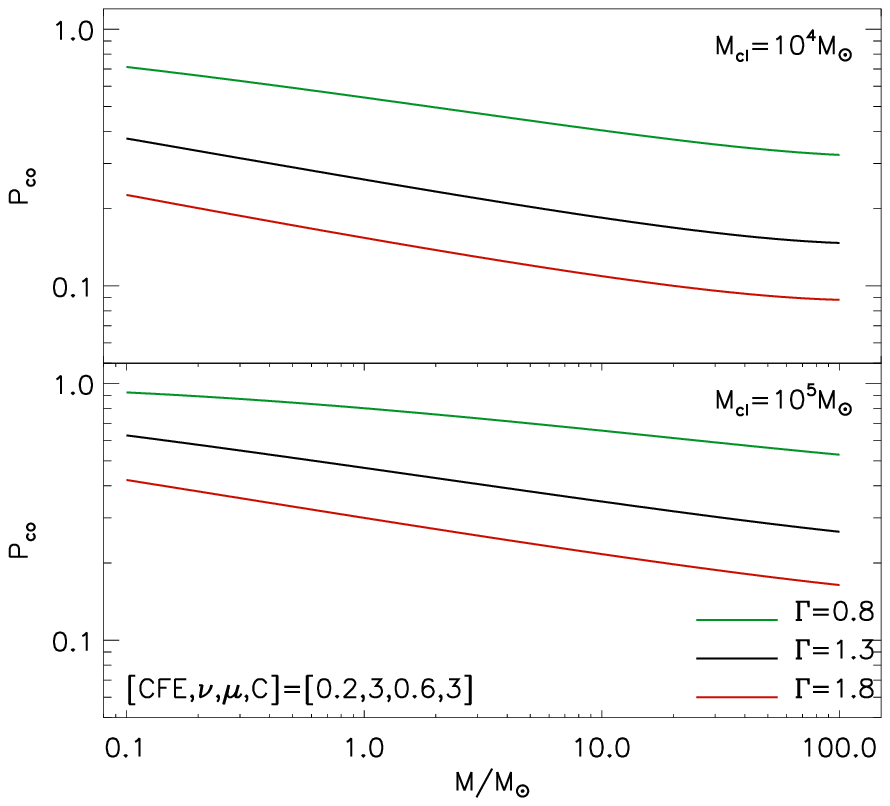}
\vskip 1cm
\caption{Coalescence probability of cores in a core mass function with fixed values of $\gamma=0.6$ and $M_{\rm ch}=1.2$ M$_{\odot}$ and for various values of $\Gamma$. The upper panel displays the case of a $10^{4}$ M$_{\odot}$ protocluster clump and the lower sub-panel  corresponds to a protocluster clump whose mass is $10^{5}$ M$_{\odot}$. In both cases, the CFE is equal to 0.2 leading to a total mass in cores of $2\times10^{3}$ M$_{\odot}$ and $2\times10^{4}$ M$_{\odot}$, respectively. The parameters of the mass-radius scaling relation are $\beta=2.5$ and $A_{\rm cl}=4000$.}
\label{fig5}
\end{center}
\end{figure}

Cores are assumed to be overdense regions with respect to their parental clump (K\"{o}nyves et al. 2020; Li et al. 2023). Observation of dense presetllar and protostellar cores in nearby molecular clouds indicate that the mean density of cores is typically a few to several times larger than the mean density of the cloud (Zhang et al. 2018; Li et al. 2023). Observations also indicate that cores that are more massive are usually more concentrated and are more dense (Johnstone \& Bally 2006). In our models, we take this into account by assuming that the density of cores can be derived from the density of the clump by

\begin{equation}
\rho_{\rm c}=C \rho_{\rm cl} \left(\frac{M}{M_{\rm l}}\right)^{\mu}
\label{eq5}
\end{equation}

where $\rho_{\rm cl}$ is the volume density of the clump which can be calculated from its mass and radius, $C$ is the over-density factor, and $M_{\rm l}$ is the minimum core mass present in the ICMF. We adopt a fiducial value of $C=3$. Equation \ref{eq5} guarantees that cores with the minimum mass have an over density of a factor $C$ with respect to the clump, whereases more massive cores are increasingly denser. Johnstone \& Bally (2006) found that, in Orion, the mass dependence of the peak density of prestellar cores is $\propto M^{0.6}$. We adopt $\mu=0.6$  as a fiducial value, albeit it is not entirely clear whether this value is representative of all environments. The lifetimes of the cores, $t_{\rm lf}$, can be expressed in units of their free fall time $t_{\rm ff}$ such that 

\begin{equation}
t_{\rm lf}=\nu t_{\rm ff}, 
\label{eq6}
\end{equation}

and, where, for spherically symmetric cores, $t_{\rm ff}$ is given by

\begin{equation}
t_{\rm ff}=\left(\frac{3 \pi}{32 G \rho_{\rm c}}\right)^{1/2},
\label{eq7}
\end{equation}

Theoretical models indicates that $\nu$ can vary between $\approx 1$ and $\approx 10$, with the latter value being the characteristic time-scale of ambipolar diffusion (McKee 1989; Tassis \& Mouschovias 2004; Das et al. 2021). Both observational (Jessop \& Ward-Thompson 2000; Kirk et al. 2005) and numerical (V\'{a}zquez-Semadeni et al. 2005; Dib et al. 2008a) estimates of gravitationally bound cores lifetimes tend to show that they are of the order of a few times their free-fall time, albeit decreasing (but still larger than one free-fall time) when cores are defined with higher density tracers. We adopt a fiducial value of $\nu=3$.

The initial radius of a core of mass $M$, $R_{0} (M)$, is calculated using its mass and density $\rho_{\rm c}$ (Eq.~\ref{eq5}). The radius of the core will decrease as time advances due to gravitational contraction and this will will lower its cross section, leading to a reduced efficiency of the coalescence process. As in Dib et al. (2007), we approximate the time evolution of the radius of the cores as: 

\begin{equation} 
R(M,t)=R_{0}(M) e^{-(t/t_{\rm lf})}.
\label{eq8}
\end{equation}

\subsection{Merger rate and probability}

When the merger of a core of mass $M$ with a core of any other mass occurs, this results in the newly formed core to be more massive than $M$. As this happens across the entire spectrum of protostellar cores masses, it leads to the formation of a larger population of more massive cores and to the depletion of lower mass ones. Even if low-mass cores are constantly replenished, this process of core coalescence, if efficient, will inevitably lead the slope of the core mass function (CMF), and subsequently of the IMF, to become shallower at the high-mass end. The efficiency of the coalescence process depends on the competition between the cores' contraction timescales and their coalescence timescales (Elmegreen \& Shadmehri 2003; Dib et al. 2007). For a core of a given mass, the higher its collision rate is, the higher is the probability for  this core to experience one or more collisions with cores of any other mass. Such a probability is given by (Hills \& Day 1976)

\begin{equation}
P_{\rm co} (M) =  \left(1-{\rm exp}\left(-\frac{t_{\rm lf}(M)}{t_{\rm co}(M)}\right)\right)
\label{eq9}
\end{equation}

where $t_{\rm lf} (M)$ is the typical core lifetime for cores of mass $M$, defined in Eq.~\ref{eq6}, and $t_{\rm co}$(M) is the typical collision timescale between cores of mass $M$ and cores of other masses. The coalescence timescale is the inverse of the coalescence rate $\omega_{\rm co}$. The coalescence rate between a core of mass $M$ and $N$ other cores of mass $M'$, $\omega_{\rm co}(M,M')$ is given by (Elmegreen \& Shadmehri 2003; Dib et al. 2007)

\begin{eqnarray}
\begin{array}{ll}
\omega(M,M')& = \frac{\sqrt{2}}{V_{\rm cl}}\sum_{j=1}^{j=N} \pi (R(M)+R(M'_{j}))^{2} v(M) \\
& \times \left[1+\frac{2 G (M+M'_{j})}{2 v^{2}(M) (R(M)+R(M'_{j}))} \right].
\end{array}
\label{eq10}
\end{eqnarray}

In Eq.~\ref{eq10}, the quantity $\pi (R(M)+R(M'))^{2}$ is the geometrical cross section, and $\left[1+\frac{2 G (M+M'_{j})}{2 v^{2}(M) (R(M)+R(M'_{j}))}\right]$ is a gravitational focusing term. The quantity $V_{\rm cl}$ is the volume of the protocluster forming clump and can be written, assuming spherical symmetry for the clump, as $V_{\rm cl}=\left(4\pi R_{\rm cl}^{3}/3\right)$. The quantity $R(M)$ is the radius of a core of mass $M$ and its initial value can be simply calculated, as mentioned above, using its mass and volume density (Eq.~\ref{eq5}). At later stages, the radius of a core of mass $M$ is calculated using Eq.~\ref{eq8} . The term $v(M)$ is the relative velocity between a core of mass $M$ and other cores. Here, we assume that the virial speed in the clump defines the relative velocity between cores of any mass, and this is given by 

\begin{equation} 
v\left(M\right)= v= \left(\frac{3 G M_{\rm cl}}{5 R_{\rm cl}}\right)^{1/2}
\label{eq11}
\end{equation}

The collision rate for a core of mass $M$ with the ensemble of cores of any mass, $\omega(M)=1/t_{\rm co}(M)$ can be simply obtained by integrating over the core mass function, $\phi_{\rm i}(M)$, such that

\begin{eqnarray}
\begin{array} {l}
\omega_{\rm co}(M) = \frac{\pi \sqrt{2} v}{V_{\rm cl}} \times \\
 \int_{M_{\rm min}}^{\inf} \phi_{\rm i} (M') (R(M)+R(M'))^{2} \left[1+\frac{2 G (M+M')}{2 v^{2} (R(M)+R(M'))}\right] dM' \\
\end{array}
\label{eq12}
\end{eqnarray}

Figures ~\ref{fig4} and ~\ref{fig5} display the ratio $\left(t_{\rm lf}/t_{\rm co}\right)$ and $P_{\rm co}$ as a function of core mass for protocluster clumps with masses of $10^{4}$ and $10^{5}$ M$_{\odot}$ (upper and lowers panels, respectively) and for a fixed set of the clumps and cores parameters, namely $\beta=2.5$, $A_{\rm cl}=4000$, and $[{\rm CFE},\nu,\mu,C]=[0.2,3,0.3,3]$. The figures display cases with $\Gamma=1.3$ (the fiducial value) and $\Gamma=0.8$ and $1.8$. The quantitative values of the ratio $(t_{\rm lf}/t_{\rm co})$ and of $P_{\rm co}$ depend on the choice of the parameters of the model. A higher value of the CFE for a given clump mass will increase the total number of cores, and hence increase the coalescence rate (i.e., smaller $t_{\rm co}$). Larger values of $\nu$ imply that the cores live longer and this will increase the ratio $(t_{\rm lf}/t_{\rm co})$. Larger values of $C$ imply that the cores are denser and thus collapse faster, leading to a decrease in the ratio $(t_{\rm lf}/t_{\rm co})$. Finally, larger values of $\mu$ imply that cores are denser than the cores of the same mass for cases with lower value of $\mu$ and thus, increasing the value of $\mu$ reduces the ratio $(t_{\rm lf}/t_{\rm co})$. The parameter $\mu$ is particularly important as its value strongly impacts the rates at which high-mass cores collapse with respect to their low mass counterparts. For the set of fiducial values of the parameters, Fig. \ref{eq4} shows that the ratio $(t_{\rm lf}/t_{\rm co})$ is smaller for higher mass cores. The ratio $(t_{\rm lf}/t_{\rm co})$ depends also on the shape of the CMF as this sets the relative numbers of high mass cores with respect to lower mass ones. 

\begin{figure*}
\begin{center}
\includegraphics[width=\textwidth] {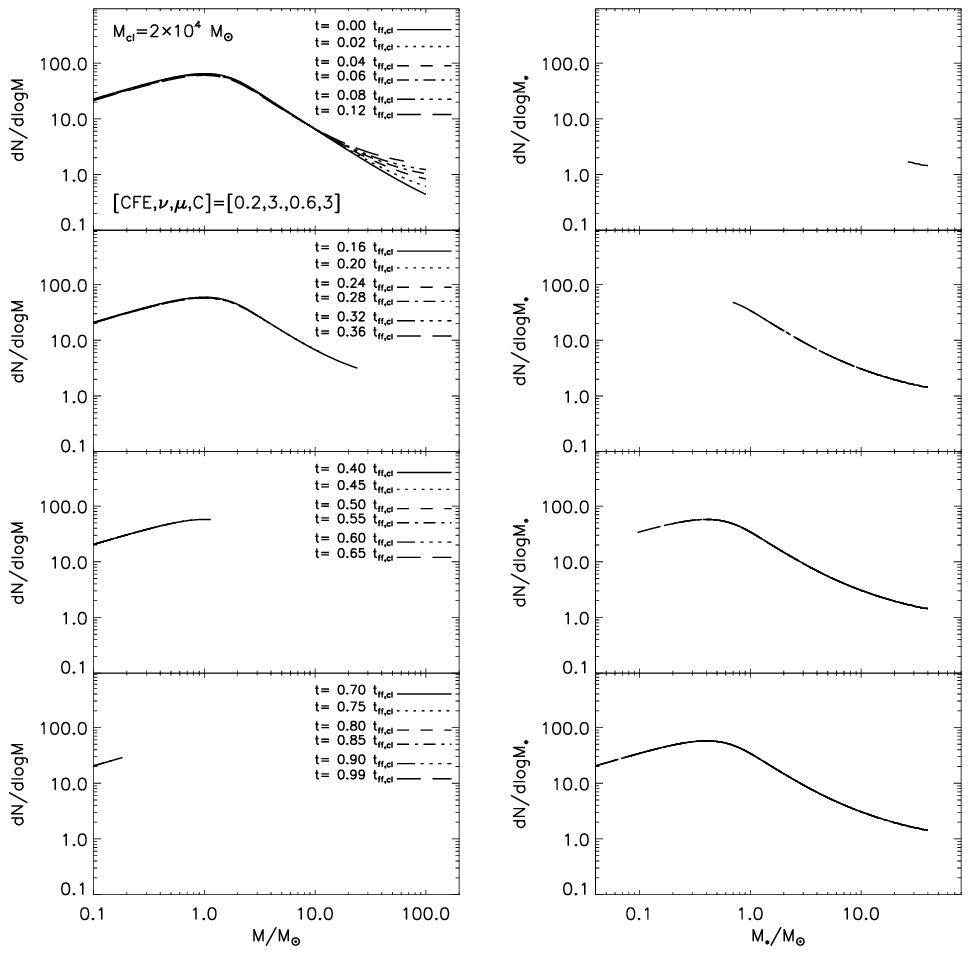}
\vskip 1cm
\caption{Time evolution of the protostellar core mass function (left) and initial stellar mass function (right) in a $2\times10^{4}$ M$_{\odot}$ mass clump and with a CFE=0.2. The parameters of the mass-radius scaling relation of the clump are $\beta=2.5$ and $A_{\rm cl}=4000$. The coalescence parameter is $\eta=0.025$. The ICMF is generated with the fiducial values of the parameters, namely $\Gamma=1.3$, M$_{\rm ch}=1.2$ M$_{\odot}$, and $\gamma=0.6$.}
\label{fig6}
\end{center}
\end{figure*}

Figure \ref{fig4} also shows that the ratio increases, at all masses, if the CMF is shallower (i.e., smaller values of $\Gamma$), as this means there is a smaller fraction of low mass cores which are, being more numerous, the most likely targets for core coalescence. This leads to a decrease of the coalescence timescale, $t_{\rm co}$, for all masses. The opposite effect is true if the CMF is steeper. An increase in the coalescence rate is also observed with increasing clump mass at at fixed value of the CFE, as larger masses increase the number of cores in the clump and reduces $t_{\rm co}$. For the same set of model parameters, Fig.~\ref{fig5} shows that the coalescence probability $P_{\rm co}$ displays the same behavior as the ratio $(t_{\rm lf}/t_{\rm co})$ with $P_{\rm co}$, being larger for more massive cores and increasing with clump mass and with a  shallower CMF. 

The ratio $(t_{\rm lf}/t_{\rm co})$ and the coalescence probability $P_{\rm co}$ can be used to compare, in a relative way, the importance of the coalescence process in clumps of different masses or with different sets of parameters. However, for any given choice of the ICMF and of the clump and cores parameters, they can not be used directly to obtain the final, post-coalescence IMF (i.e., after all cores have collapsed and turned into main sequence stars). This is simply because the coalescence process is highly non-linear and the number of cores of a given mass is continuously changing over time with lower mass cores being depleted and higher mass cores being created while at the time cores are removed from the time-dependent core mass function (CMF) and populating the IMF as they continuously collapse to form stars. Below, we describe this process of the co-evolution of the CMF and the IMF. 

\section{The co-evolution of the CMF and of the IMF}\label{coevol}

In order to follow the evolution of a population of coalescing cores in a protocluster forming clump, it is necessary to solve the following time-dependent equation

\begin{eqnarray} 
\frac{dN(M,t)}{dt}=0.5\times \eta \times \nonumber \\
 \int^{\Delta M}_{M_{min}}~N(m,t)~N(M-m,t)~\sigma(m,M-m)~v~dm \nonumber \\ 
 -\eta  N(M,t) \int^{M_{max}}_{M_{min}} N(m,t) \sigma(m,M) v~dm,                            
\label{eq13}
\end{eqnarray}

where the first and second terms in the right hand side of Eq.~\ref{eq13} correspond to the rate of creation and destruction of cores of mass $M$, respectively. In Eq.~\ref{eq13}, $\Delta M=M-M_{min}$, and $\eta$ is a coefficient which represents the coalescence efficiency (i.e., the probability for two cores to merge), with $\eta \leq 1 $. This efficiency parameter encapsulates the effects of various physical processes that can affect the coalescence of cores, such as if the encounter of cores occurs preferentially parallel or perpendicular to the local magnetic field lines. In all generality, $\eta$ may depend on the mass of the cores and their position in the clump, and may also vary with time. As we are considering uniform density clumps, $\eta$ is independent of location, and we further posit that it does not depend on the core mass and does not vary with time. We consider two values of $\eta$ of 0.01 and 0.025. Cores are injected into the clump in a single burst with a mass distribution that follows the ICMF and a normalization that is dictated by the clump mass and the chosen CFE. In order to evaluate when cores of mass $M$ have to be removed from the CMF and populate the IMF as newly formed stars, the time elapsed since the formation of the cores is compared to the cores lifetimes, $t_{\rm lf}$. Cores of a given mass whose lifetimes become smaller than the elapsed time are removed from the CMF and transferred to the IMF. When cores are turned into stars, only a fraction of the core mass will end up being locked in the star, such that the mass of the star is given by $M_{*}=\epsilon_{*} M$, with $\epsilon_{*}$ being the core-to-star conversion efficiency with $\epsilon_{*} \leq 1$. In this work, we do not include a sub-fragmentation model and we count a single star or multiple system resulting from the collapse of cores as one object. This approach is valid since we will be comparing our models to observations (data in Fig.~\ref{fig2}) for which no multiplicity correction has been applied. Matzner \& McKee (2000) found that $\epsilon_{*}$ can vary between 0.25 and 0.7 for stars in the mass range [0.5-2] M$_{\odot}$. To date, there is no clear evidence whether this result holds at higher core masses. In this work, we adopt an intermediate value of $\epsilon_{*}=0.4$ as a fiducial value and assume that it is independent of core mass.  

\begin{figure}
\begin{center}
\includegraphics[width=\columnwidth] {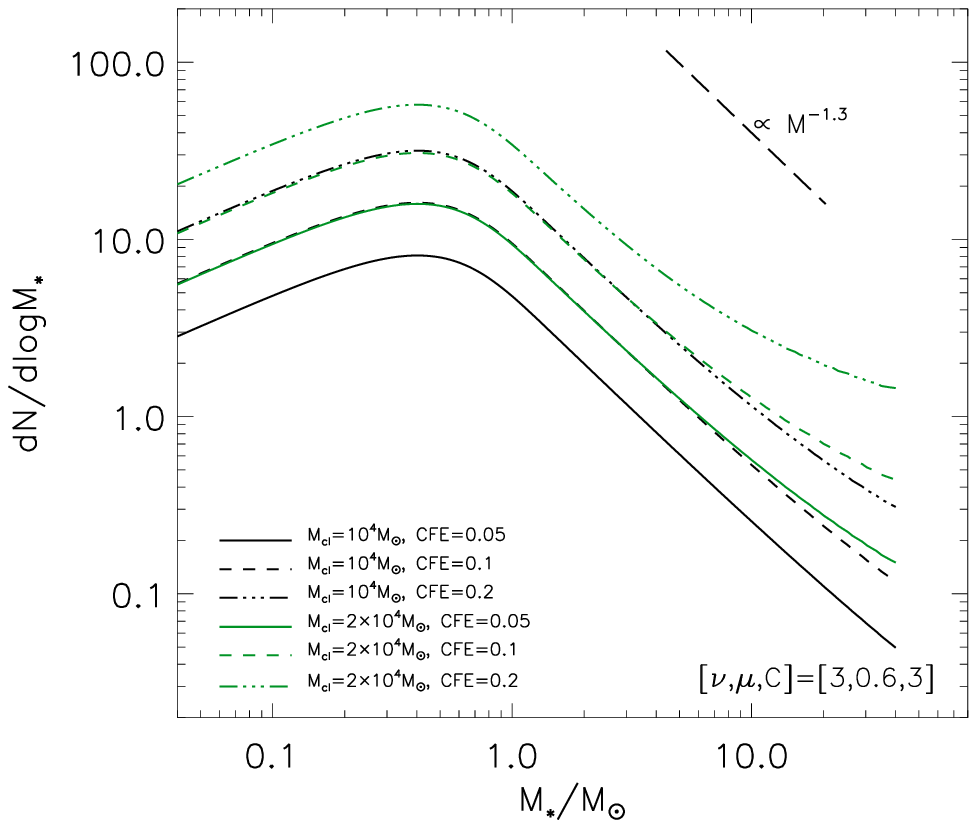}
\vskip 1cm
\caption{Final, post-coalescence IMF for clumps with masses $M_{\rm cl}=10^{4}$ and $2\times10^{4}$ M$_{\odot}$ and a CFE of 0.05, 0.1, and 0.2. The parameters of the mass-radius scaling relation of the clump are $\beta=2.5$ and $A_{\rm cl}=4000$. The coalescence parameter is $\eta=0.025$. In all models, the ICMF is generated with the fiducial values of the parameters, namely $\Gamma=1.3$, M$_{\rm ch}=1.2$ M$_{\odot}$, and $\gamma=0.6$.}
\label{fig7}
\end{center}
\end{figure}

\begin{figure*}
\centering
\includegraphics[width=0.32\textwidth]{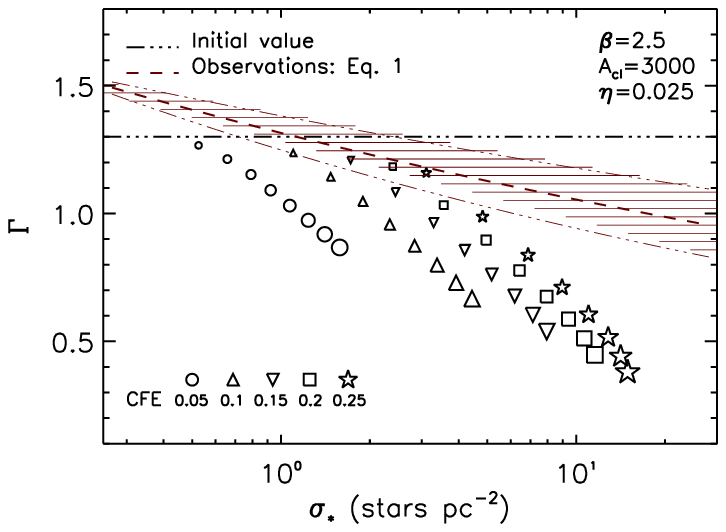}
\includegraphics[width=0.32\textwidth]{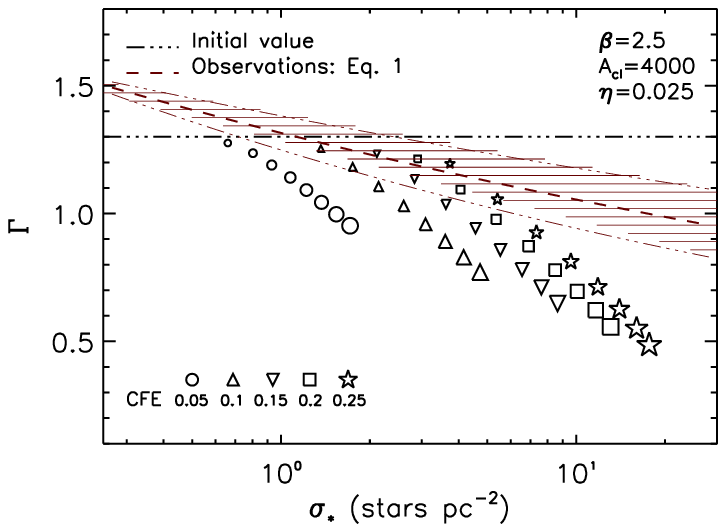}
\includegraphics[width=0.32\textwidth]{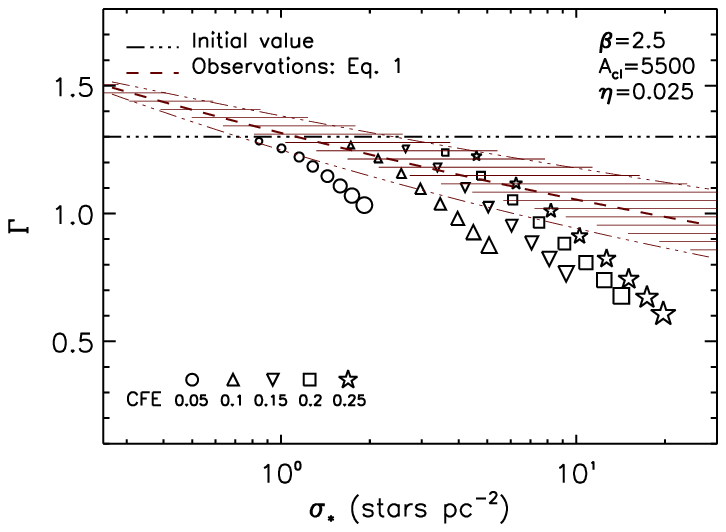}\\
\includegraphics[width=0.32\textwidth]{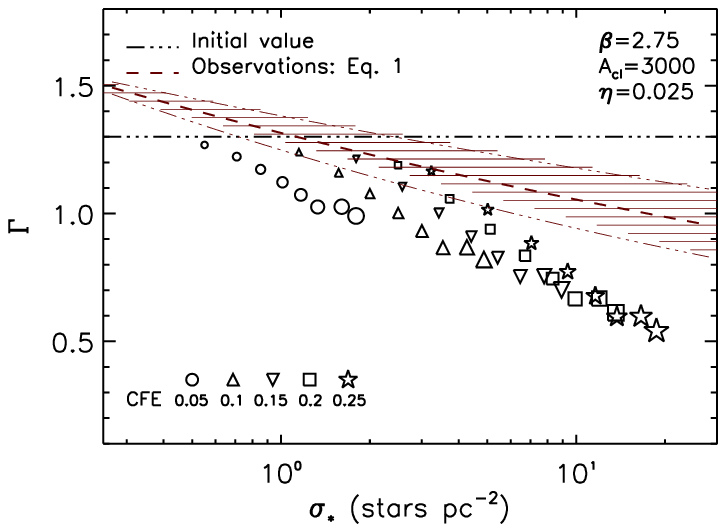}
\includegraphics[width=0.32\textwidth]{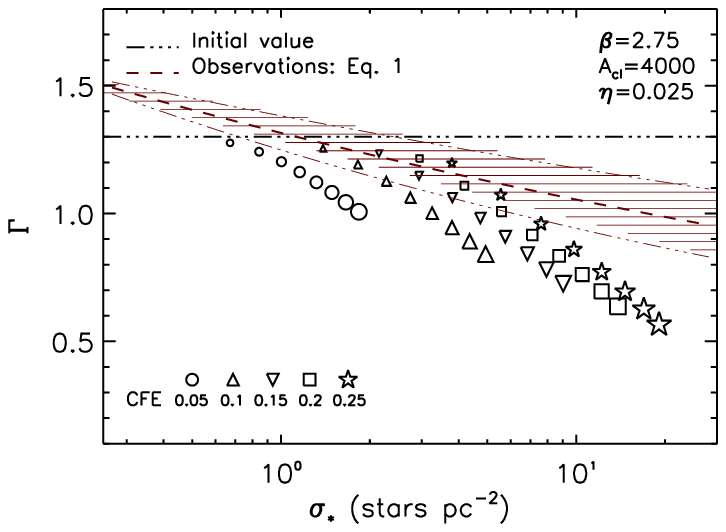}
\includegraphics[width=0.32\textwidth]{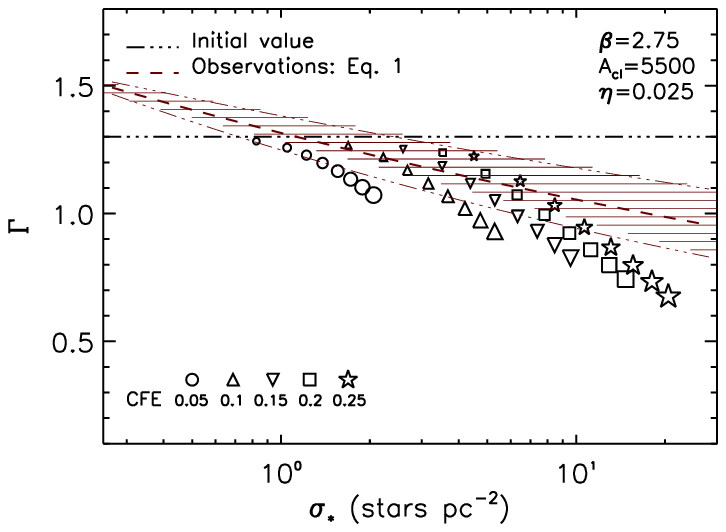}\\
\includegraphics[width=0.32\textwidth]{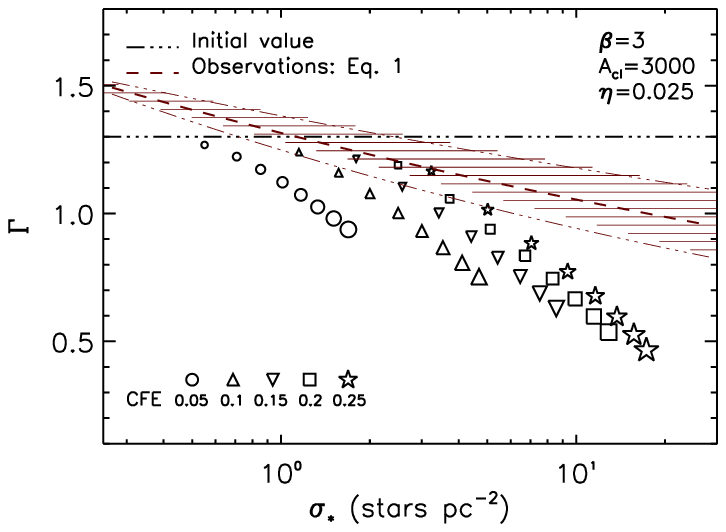}
\includegraphics[width=0.32\textwidth]{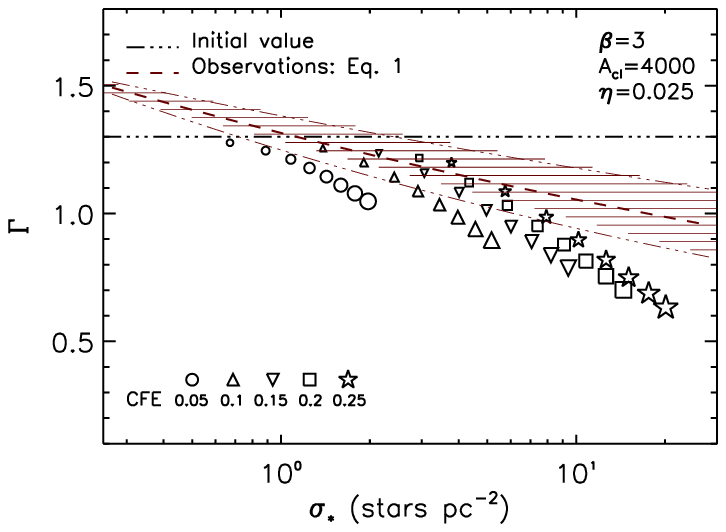}
\includegraphics[width=0.32\textwidth]{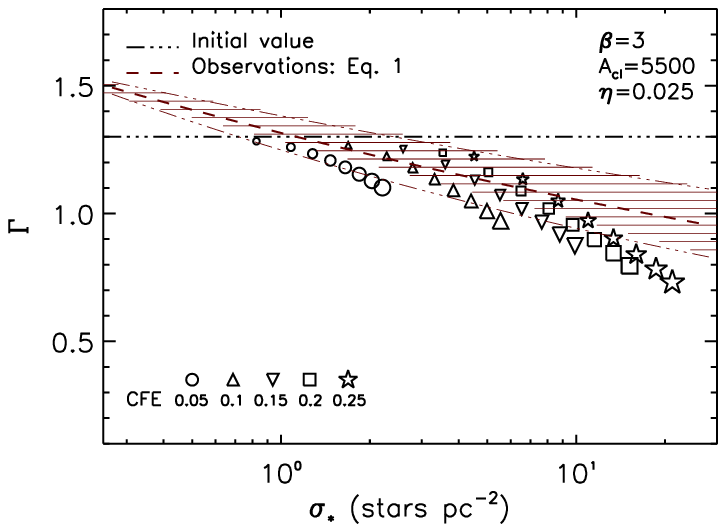}
\caption{The value of the slope of the post-coalescence IMF at the high-mass end (mass range [1-40] M$_{\odot}$) in the models as a function of the stellar surface density. The size of the symbols reflects the mass of the protocluster clumps starting from $5\times10^{3}$ M$_{\odot}$ for the smallest symbols size up to $4\times10^{4}$ M$_{\odot}$ increasing by steps of $5\times10^{3}$ M$_{\odot}$. The value of the coalescence efficiency parameter is fixed to $\eta=0.025$ in all models and the core-to-star efficiency parameter is $\epsilon_{*}=0.4$. The purple line is a fit to the observations (Eq.~\ref{eq1}). The dashed dotted line is the initial value of the slope for the ICMF in all models.}
\label{fig8}
\end{figure*}

\begin{figure*}
\centering
\includegraphics[width=0.32\textwidth]{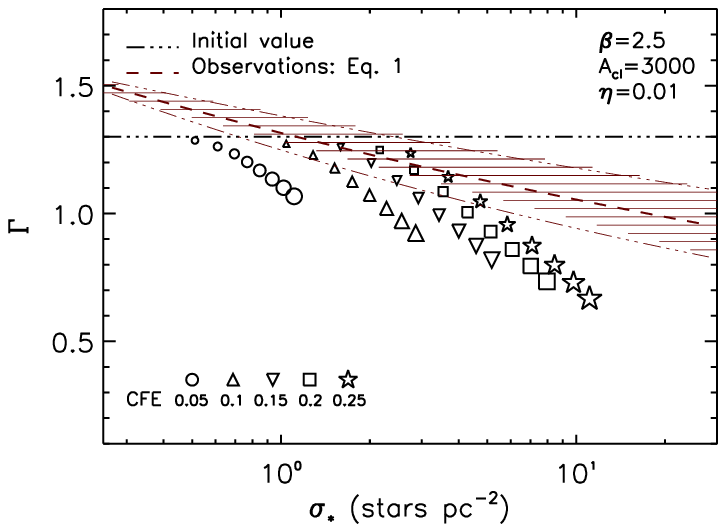}
\includegraphics[width=0.32\textwidth]{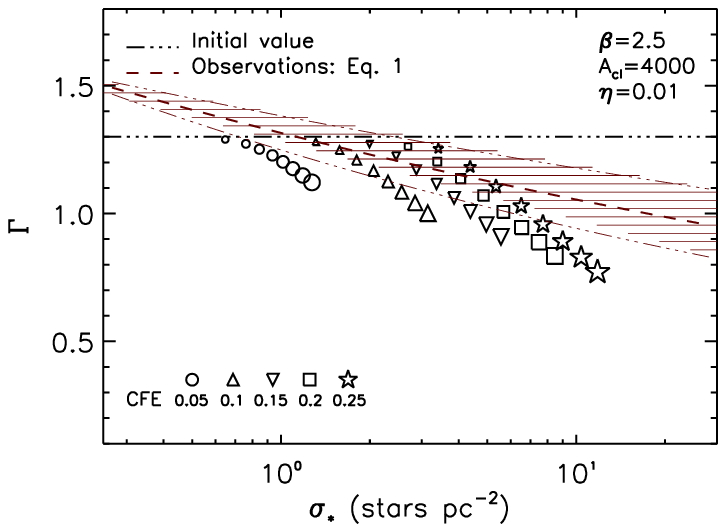}
\includegraphics[width=0.32\textwidth]{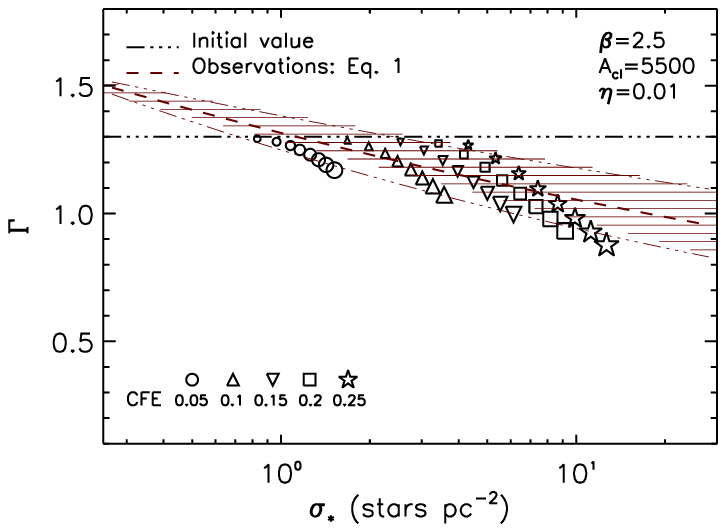}\\
\includegraphics[width=0.32\textwidth]{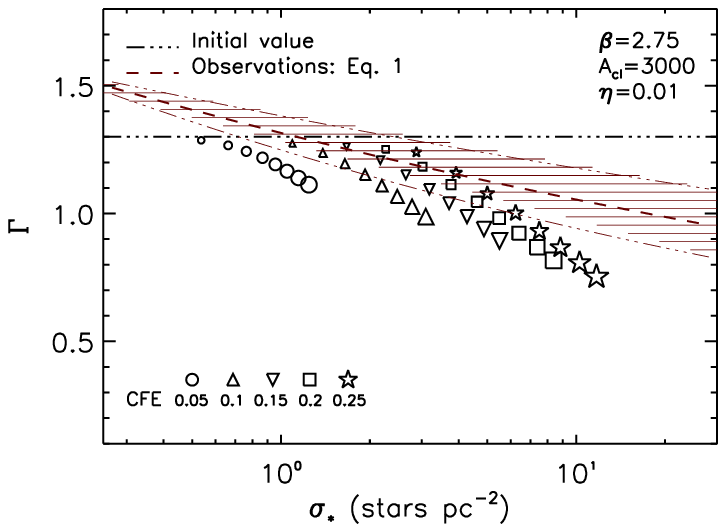}
\includegraphics[width=0.32\textwidth]{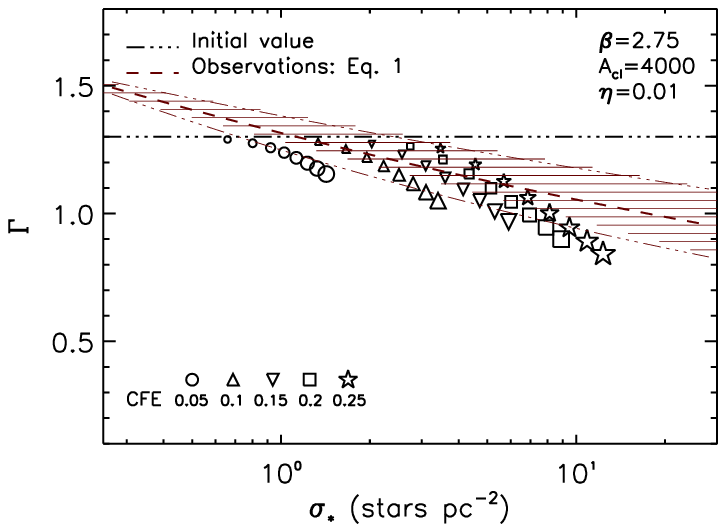}
\includegraphics[width=0.32\textwidth]{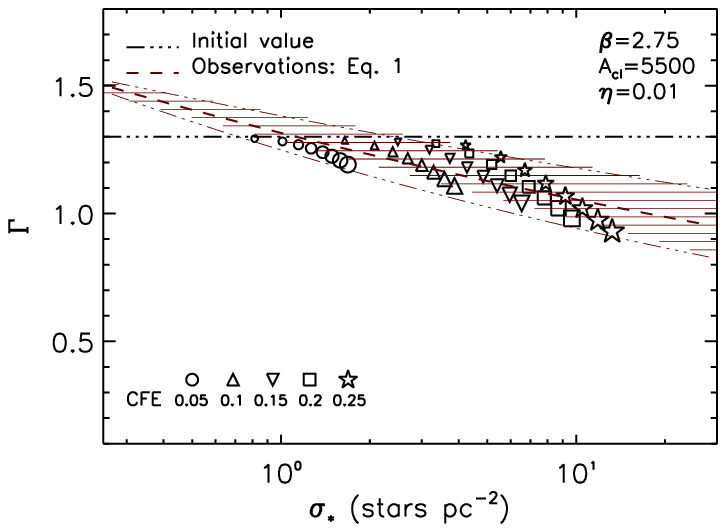}\\
\includegraphics[width=0.32\textwidth]{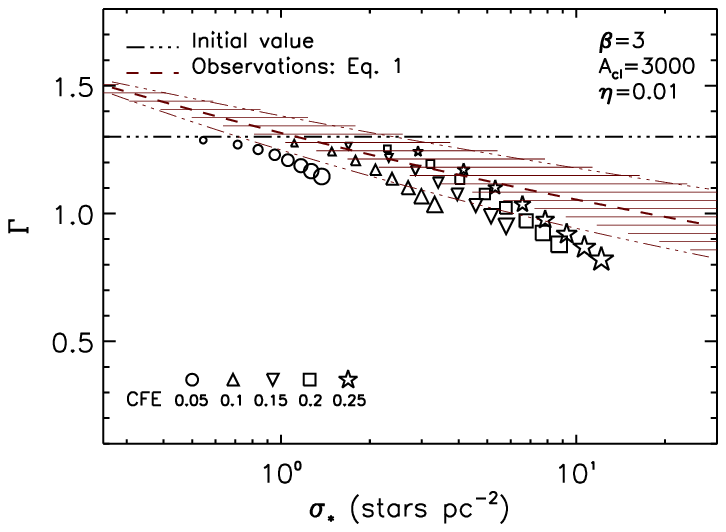}
\includegraphics[width=0.32\textwidth]{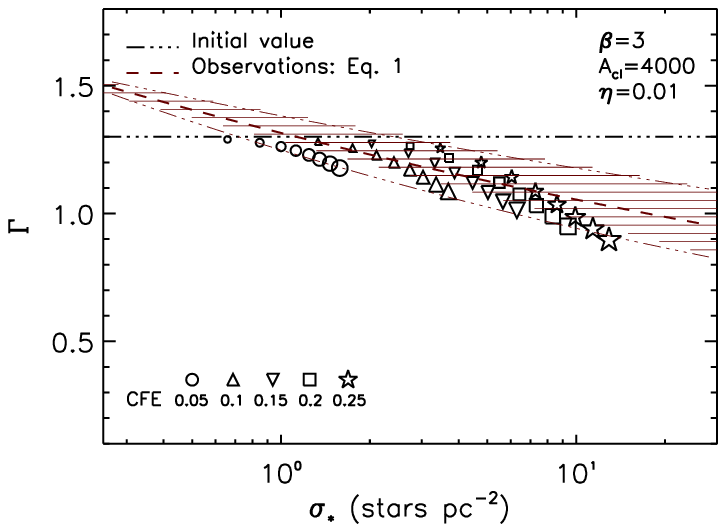}
\includegraphics[width=0.32\textwidth]{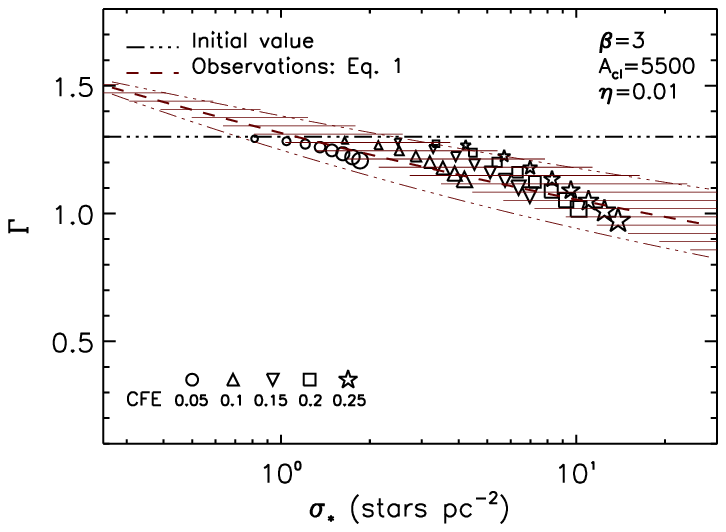}
\caption{The value of the slope of the post-coalescence IMF at the high-mass end (mass range [1-40] M$_{\odot}$) in the models as a function of the stellar surface density. The size of the symbols reflects the mass of the protocluster clumps starting from $5\times10^{3}$ M$_{\odot}$ for the smallest symbols size up to $4\times10^{4}$ M$_{\odot}$ increasing by steps of $5\times10^{3}$ M$_{\odot}$. The value of the coalescence efficiency parameter is fixed to $\eta=0.01$ in all models and the core-to-star efficiency parameter is $\epsilon_{*}=0.4$. The purple line is a fit to the observations (Eq.~\ref{eq1}). The dashed dotted line is the initial value of the slope for the ICMF in all models.}
\label{fig9}
\end{figure*}

As an example, Fig.~\ref{fig6} displays the time evolution of the CMF and of the IMF in a protocluster clump with a mass $M_{\rm cl}=2\times10^{4}$ M$_{\odot}$ and that obeys a mass-radius relation with $\beta=2.5$ and $A_{\rm cl}=4000$. The other parameters of the model are also set to their fiducial values, namely $[{\rm CFE},\nu,\mu,C]=[0.2,3,0.6,3]$ and the coalescence efficiency is set to $\eta=0.025$. Time in Fig.~\ref{fig6} is displayed in units of the clump's free fall time $t_{\rm ff,cl}$, which, for this particular model, is $t_{\rm ff,cl}\approx 3\times10^{5}$ yr. The left-hand panel of Fig.~\ref{fig6} display the time evolution of the CMF and the right-hand panel, that of the IMF. Initially, there are no stars in the clump, and cores are injected in a single burst with an ICMF whose parameters are the fiducial ones ($\Gamma=1.3$, M$_{\rm ch}=1.2$ M$_{\odot}$, $\gamma=0.6$; full line in the upper left sub panel). What is observed is that a rapid flattening occurs at the high mass end of  the CMF, and is accompanied by a decline in the numbers of low mass cores as these are the most likely targets for coalescence, being more numerous, and also slower to collapse. As cores start to collapse, starting with the most massive ones, they are moved to the IMF which retains the flattening of the CMF at the high-mass end. In order to further illustrate the effects of some of the model's parameters, we compare in Fig.~ \ref{fig7} the final, post-coalescence IMF, for clumps of mass $M_{\rm cl}=10^{4}$ M$_{\odot}$ and $M_{\rm cl}=2\times10^{4}$ M$_{\odot}$ and for a CFE of 0.05, 0.1, and 0.2. In all of these models, $\eta=0.025$, and the ICMF is generated with the fiducial values of the parameters. Figure \ref{fig7} clearly indicates that the IMF at the high-mass end is systematically shallower for higher masses of the protocluster clump and for higher values of the CFE. 

In order to enable a comparison between the models and the observations (i.e., Fig.~\ref{fig2}), we have to derive the values of $\Gamma$ and $\sigma_{*}$ in the models in a similar way to the observations. We calculate the value of $\Gamma$ by fitting the post-coalescence IMF with a power-law for stars in the mass interval [$1,\epsilon_{*}\times M_{\rm u}$] M$_{\odot}$, which for our choice of $\epsilon_{*}=0.4$ and $M_{\rm u}=100$ M$_{\odot}$, corresponds to an upper limit on the stellar mass of $40$ M$_{\odot}$. For $\sigma_{*}$, and adopting the same approach as in the observations (Lim et al. 2015a), we estimate its value using stars more massive than $5$ M$_{\odot}$ ($N_{*} (M \geq 5$ M$_{\odot}$)) such that

\begin{equation}
\sigma_{*} = \frac {N_{*} (M \geq 5 {\rm M_{\odot}})} {\pi R_{\rm cl}^{2}},
\label{eq14}
\end{equation}  

where $R_{\rm cl}$ is the clump's radius and N$_{*}$ ($M \geq 5$ M$_{\odot}$) is given by

\begin{equation}
N_{*} (M \geq 5 {\rm M_{\odot}})=\int_{5 {\rm M_{\odot}}}^{40 M_{\odot}} \phi_{\rm pc}(M_{*}) dM_{*}.
\label{eq15}
\end{equation}

In Eq.~\ref{eq15}, $\phi_{\rm pc}$, represents the final, post-coalescence IMF. The choice of $\epsilon_{*}$ does not affect the derived value of $\Gamma$ if $\epsilon_{*}$ is taken to be a constant, independent of the core mass. However, the value of $\epsilon_{*}$ does affect the derived value of $\sigma_{*}$ and thus the position of the model in the $\Gamma$ vs. $\sigma_{*}$ plot. However, we have tested that variations of $0.1$ around our fiducial value of $\epsilon_{*}=0.4$ have no significant influence on the position of any given model in the $\Gamma$-$\sigma_{*}$ space. 

\section{Parameter study and comparison to the observations}\label{paramstudy}

\begin{figure}
\centering
\includegraphics[width=0.9\columnwidth]{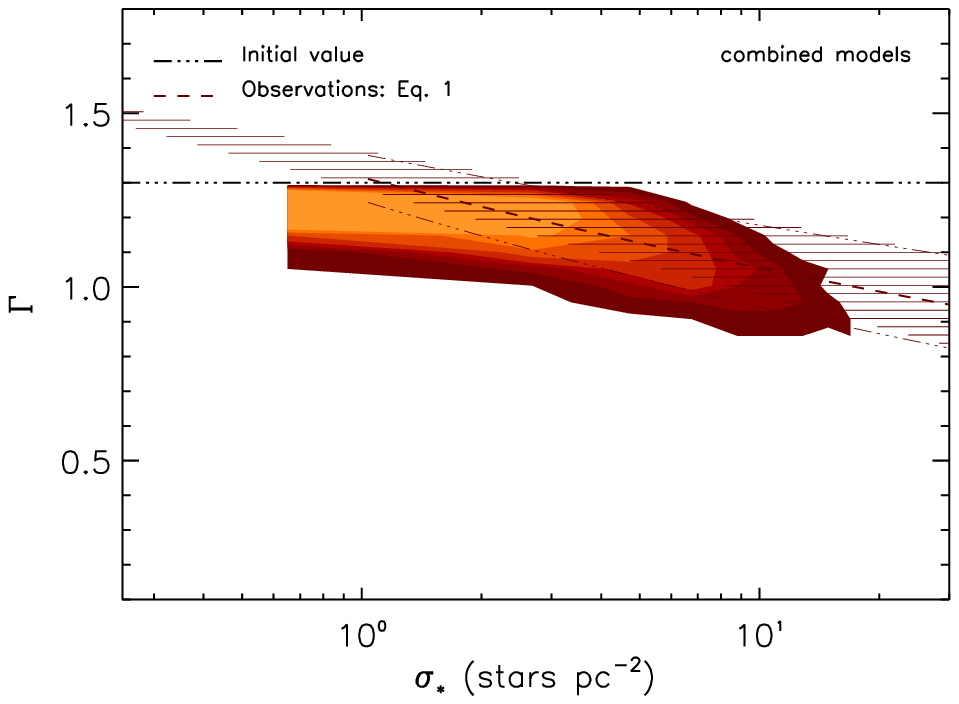}
\vskip 1cm
\caption{The $\Gamma$ vs. $\sigma_{*}$ relation based on combining all models that have $A_{\rm cl}=5500$ and $\eta=0.01$ and for the three values of $\beta$ ($2.5, 2.75$ and $3$), and 4 values of the core-to-star-efficiency parameter $\epsilon_{*} (0.35,0.4,0.45$ and $0.5$). The models are displayed as a 2D histogram and compared to the fit to the observations (Eq.~\ref{eq1}). The shaded area represents the 1$\sigma$ uncertainty of the observations. The color contours, going from dark colors to lighter colors correspond to 1, 4, 7, 10, 15, 20 and 25 occurrences, respectively.}
\label{fig10}
\end{figure}

We explore the parameter space by varying the coefficient and exponent of the mass-radius relation of protocluster clumps. As stated above, we consider three value of $\beta=2.5$, $2.75$ and $3$ as well as three values of $A_{\rm cl}=3000$, $4000$, and $5500$. We also consider protocluster clump masses that vary between $5\times10^{3}$ M$_{\odot}$ and $4\times10^{4}$ M$_{\odot}$ and values of the CFE that are between 0.05 and 0.25. In all of these models, the parameters of the cores are fixed to $[\nu,\mu,C]=[3,0.6,3]$. We also consider two values of the coalescence efficiency parameter of $\eta=0.01$ and $\eta=0.025$. For each model, we measure the value of $\Gamma$ in the post-coalescence IMF and estimate $\sigma_{*}$ using Eq.~\ref{eq14}. Figure \ref{fig8} displays the result in the $\Gamma$-$\sigma_{*}$ space for a number of models with $\eta=0.025$ and for various permutations of $\beta$ and $A_{\rm cl}$. Each sub-panel contains models with different values of the mass of the clump, $M_{\rm cl}$ and of the CFE. Figure \ref{fig9} is similar to Fig.~\ref{fig8} but it pertains to models where the coalescence process is less efficient (i.e., $\eta=0.01$). In both cases, the models are compared to the fit to the observational data (Eq.~\ref{eq1}) in the stellar surface density regime covered by the models. A number of things can be learned from the results displayed in Figs.~\ref{fig8} and \ref{fig9}. For clump models where the total mass in cores formed in the burst is small (i.e., for small clump masses and small CFE), the flattening of the slope with respect to the original value of the ICMF ($\Gamma=1.3$) is small and would not be detected in the observations within the $1\sigma$ uncertainty. However, in all models, the value of $\Gamma$ decreases with increasing stellar surface density. The figures also show that some models reproduce the observations better than others. Models with $\eta=0.025$ (Fig.~\ref{fig8}) tend to produce shallower slopes at the corresponding stellar surface density, whereas models with $\eta=0.01$ are in much better agreement with the observations. Furthermore, in both figures and particularly for the case with $\eta=0.01$, the observations are better reproduced with models of more compact clumps that have larger values of $\beta$ and $A_{\rm cl}$. In these models, clumps have a smaller radius $R_{\rm cl}$, for a given mass $M_{\rm cl}$. 

For the best fitting models, that is for those with $\eta=0.01$ and $A_{\rm}=5500$ and for all values of $\beta$ ($2.5$, $2.75$, and $3$), we have generated additional realizations in which we have varied the value of $\epsilon_{*}$. In addition to the fiducial value of $\epsilon_{*}=0.4$, we consider values of $\epsilon_{*}=0.35$, $0.45$, and $0.5$. All of these models are combined together and displayed as a 2D histogram in Fig.~\ref{fig10}. The ensemble of these models is compared to the power law fit of the observations that are displayed in Fig.~\ref{fig2} (i.e., Eq.~\ref{eq1}). What Fig.~\ref{fig10} shows is that this set of models reproduces the observations remarkably well. This indicates that the coalescence of protostellar cores in protocluster forming clumps is not only a viable channel for producing IMFs that are shallower than the Salpeter mass function but for relating the slope of the IMF of young clusters to some of their fundamental properties. 

\section{Discussion and conclusions}\label{conc}

In this work, we compiled a number of observational results of the high-mass slope of the IMF of young Galactic stellar clusters and of the slope of the protostellar core mass function of a massive Galactic star forming region. We found that the exponent of the power law that describes the shape of the IMF in this mass regime ($\Gamma$) is related to the clusters' stellar surface density ($\sigma_{*}$) with a dependence that follows $\Gamma=1.31~\sigma_{*}^{-0.095}$. 

In order to explain these observational findings, we investigated the effects of the coalescence of dense protostellar cores in protocluster forming clumps on the shape of the IMF in the regime of intermediate to high-mass stars. We used a simple model in which clumps are modeled as uniform density spheres whose radii are linked to their mass via a parametrized mass-radius relation (M$_{\rm cl}=A_{\rm cl}\times R_{\rm cl}^{\beta}$) and have a burst-like formation of dense cores with a given core formation efficiency (CFE). Protostellar cores are modeled as uniform spheres that are over-dense with respect to the clump by at least a factor $C$ (for the least massive) and the most massive cores are even denser due to an imposed mass-overdensity relation. Cores have finite lifetimes that are parametrized as a number of times their free-fall time. Given a CFE of the clump, cores are introduced as a single burst with a prescribed core mass function (ICMF), typically one that has a slope at the high-mass end similar to the Salpeter value. The coalescence of the cores is followed over time and when their lifetimes becomes smaller than the elapsed time since they were formed, they are removed from the time evolving core mass function and transferred to the IMF. 

Our models indicate that the occurrence of coalescence generates a flattening of the slope of the IMF in the intermediate to high-mass end. The level of flattening depends on the coalescence efficiency (parameter $\eta$ in the text) which describes the ability of two cores of mass $M$ and $M'$ to merge and form a core of mass $M+M'$. It also depends on the clump properties. We find that the observational $\Gamma-\sigma_{*}$ relation is best reproduced by models in which clumps are compact (i.e., large values of $\beta$ and $A_{\rm cl}$). In future work, we will investigate more realistic configurations with clumps having a radial density profile and different levels of mass segregation as those observed in Galactic star forming regions (Alfaro \& Rom\'{a}n-Z\'{u}\~{n}iga 2018; Dib \& Henning 2019; Nony et al. 2023). We also plan to explore the effects of a time-dependent core formation history in the clumps. 

Accounting for, and understanding the origin of IMF variations in clusters and the connection between the level of variations and the properties of young clusters has important consequences for our understanding of galaxy evolution, both for the Milky Way and external galaxies. Several authors have already pointed out the need to account for IMF variations. Guszejnov et al. (2019) showed, using numerical simulations of both Early type galaxies as well as of Ultra Faint Dwarf galaxies, that the present day mass function of these galaxies which is observed to be bottom heavy and top heavy, respectively, cannot be reconciled with the concept of a universal IMF in clusters. As Guszejnov et al. (2019) stated, either the present day mass function of these galaxies is problematic or the universality of the IMF in clusters is in question. Along this same idea, Dib \& Basu (2018) and Dib (2022) showed how the presence of significant cluster-to-cluster of the IMF can generate a bottom heavy mass function in Elliptical galaxies. Additionally, Dib (2022) showed that accounting for IMF variations in clusters coupled with a metallicity dependence of the IMF, can help reproduce the galaxy-wide IMF of Ultra Faint Dwarf galaxies. 

\begin{acknowledgments}

I thank Beomdu Lim and Matthew Hosek for providing valuable insight into the Sejong survey of stellar clusters and the Arches cluster, respectively. I am also grateful to Mohsen Shadmehri, Valery Kravtsov, and S\'{e}bastien Comer\'{o}n for sharing useful comments on an earlier version of this paper. I also thank the anonymous referee for useful comments. 
 
\end{acknowledgments}

%\appendix
%\section{Appendix information}

%\bibliography{sample631}{}
%\bibliographystyle{aasjournal}
{}

\end{document}